# Synthesis and Properties of *Non-Curing* Graphene Thermal Interface Materials

Sahar Naghibi[1,2], Fariborz Kargar[1,3,*], Dylan Wright[1,2], Chun Yu Tammy Huang[1,2], Amirmahdi Mohammadzadeh[1,3], Zahra Barani[1,3], Ruben Salgado[1,2], and Alexander Balandin[1,2,3,*]

[1]Phonon Optimized Engineered Materials (POEM) Center, University of California, Riverside, California 92521 USA

[2]Materials Science and Engineering Program, Bourns College of Engineering, University of California, Riverside 92521 USA

[3]Department of Electrical and Computer Engineering, Bourns College of Engineering, University of California, Riverside, California 92521 USA

* Corresponding authors: fkargar@ece.ucr.edu and balandin@ece.ucr.edu ; web-site: http://balandingroup.ucr.edu/

**Abstract**

Development of the next generation thermal interface materials with high thermal conductivity is important for thermal management and packaging of electronic devices. We report on the synthesis and thermal conductivity measurements of *non-curing* thermal paste, *i.e.* grease, based on mineral oil with the mixture of graphene and few-layer graphene flakes as the fillers. It was found that graphene thermal paste exhibits a distinctive thermal percolation threshold with the thermal conductivity revealing a sublinear dependence on the filler loading. This behavior contrasts with the thermal conductivity of *curing* graphene thermal interface materials, based on epoxy, where super-linear dependence on the filler loading is observed. The performance of graphene thermal paste was benchmarked against top-of-the-line commercial thermal pastes. The obtained results show that non-curing graphene thermal interface materials outperforms the best commercial pastes in terms of thermal conductivity, at substantially lower filler concentration of $f_g = 27$ vol%. The obtained results shed light on thermal percolation mechanism in non-curing polymeric matrices laden with quasi-two-dimensional fillers. Considering recent progress in graphene production via liquid phase exfoliation and oxide reduction, we argue that our results open a pathway for large-scale industrial application of graphene in thermal management of electronics.

## I. Introduction

As transistors continue to decrease in size and packing densities increase, thermal management becomes the critical bottleneck for development of next generation of compact and flexible electronics[1]. The increase in computer usage and ever-growing dependence on cloud systems require better methods for dissipating heat away from electronic components. The important ingredients of thermal management are the thermal interface materials (TIMs). Various TIMs interface two uneven solid surfaces where air would be a poor conductor of heat, and aid in heat transfer from one medium into another. Two important classes of TIMs include curing and non-curing composites. Both of them consist of a base, *i.e.* matrix materials, and thermally conducting fillers. Commonly, the studies of new fillers for the use in TIMs start with the curing epoxy-based composites owing to the relative ease of preparation and possibility of comparison with a wide range of other epoxy composites. Recent work on TIMs with carbon fillers have focused on curing composites, which dry to solid[2–7]. Curing TIMs are required for many applications, *e.g.* attachment of microwave devices, but do not cover all thermal management needs. Thermal management of computers requires specifically non-curing TIMs, which are commonly referred to as thermal pastes or thermal greases. They are soft pliable materials, which unlike cured epoxy-based composites, or phase change materials, remain soft once applied. This aids in the avoiding crack formations in the bond line due to repeated thermal cycling of two connected materials with different temperature expansion coefficients. Non-curing TIMs also allow for easy reapplication, known as a TIM's re-workability property. Non-curing TIMs are typically cost efficient – an essential requirement for commercial applications. Various applications in electronics, non-curing grease-like (soft) TIMs are preferred. Examples of the applications include but are not limited to cooling of large data centers [8] and personal devices which are the primary targets for these applications. Current commercially available TIMs perform in thermal conductivity range of 0.5 $Wm^{-1}K^{-1}$ to 5 $Wm^{-1}K^{-1}$ with combination of several fillers at high loading fractions [ref. [9]]. State−of−the−art and next generation electronic devices require thermal pastes with bulk thermal conductivity in the range of 20 to 25 $Wm^{-1}K^{-1}$ [10,11]This study focuses specifically on non-curing TIMs with graphene and few-layer graphene fillers.

Curing and non-curing TIMs consists of two main components – a polymer or oil material as its base and fillers, which are thermally conductive inclusions added to the base increasing the overall heat conduction properties of the resulting composite. Polymer base materials have a rather low thermal conductivity within the range of 0.2 $Wm^{-1}K^{-1}$ to 0.5 $Wm^{-1}K^{-1}$, mainly owing to their amorphous structure[12]. The strategy for creating advanced TIM is to find a filler with high intrinsic thermal conductivity and incorporate it into a base creating a soft material, which is easy to apply and bind the interfaces. Numerous other parameters such as filler – matrix coupling, uniformity of the dispersion of the fillers, viscosity, and surface adhesion affect the resulting performance of the TIM. Conventional fillers, which are added to enhance the thermal properties of the base polymeric or oil matrices, span a wide range of materials, including metals[13,14], ceramics, metal oxides[15–20], and semiconductors[18,21] with micro and nanometer scale dimensions. Apart from thermal conductivity, the selection criteria for fillers include many parameters such as compatibility with the matrix, weight, thermal expansion characteristics and rheological behavior. Recent concerns over environmentally friendly materials further limit the list of available additives, which can be used as fillers. Considering all these parameters and limitations, the most promising recently emerged filler material is graphene[22,23].

The first exfoliation of graphene[24,25] and measurement of its electrical properties sparked intensive efforts to find graphene's applications in electronics[26], *e.g.* as on-chip or inter-chip[27] interconnects[28,29], or a complementary material to silicon in analog or non-Boolean electronics[30]. The idea of using graphene as fillers in thermal applications emerged from the discovery of the exceptional heat conduction properties of suspended "large" flakes of single layer graphene (SLG), with the thermal conductivity ranging from 2000 $Wm^{-1}K^{-1}$ to 5300 $Wm^{-1}K^{-1}$.[31,32] It is established that acoustic phonons are the main heat carriers with the "gray" mean-free-path (MFP) of ~750 nm. Theory suggests that long-wavelength phonons with much larger MFP make substantial contribution to thermal conductivity. The thermal conductivity of SLG with lateral dimensions smaller than MFP degrades due to the "classical size" effects, *i.e.* phonon – flake edges scattering. The thermal conductivity of SLG is vulnerable to defects, wrinkles, bending, and rolling[33]. The cross-section of SLG is also small making it not an ideas filler. From another perspective, FLG is more resistant to degradation of its intrinsic thermal properties due to rolling, bending or exposure to matrix defects. For these reasons, FLG with some addition of SLG, create better filler-matrix and filler-filler coupling, and are considered to be optimum filler mixture. The in-plane thermal conductivity of FLG converges to that of the high quality bulk graphite, which by itself is as high as ~2000 $Wm^{-1}K^{-1}$, as the number of layers exceeds about eight mono-layers[34–36]. The ability of FLG – a van der Waals material – to present thermal conductivity of bulk graphite is an important factor for thermal applications. The thermal conductivity of FLG is one and two orders of magnitudes higher than that of the conventional metallic and ceramics fillers, respectively.

Technological challenges using graphene and FLG as fillers in thermal management applications, which by their nature requires large amount of source material, were linked to the low yield production laboratory methods. The last decade of graphene research has led to development of several scalable techniques, such as liquid phase exfoliation (LPE)[37,38] and graphene oxide reduction[39,40], which provide large quantities of graphene and FLG of quality acceptable for thermal applications, making the mass production cost effective[38,41]. These recent developments remove the barriers for graphene utilization in the next generation of curing and non-curing TIMs. In the following discussion, in thermal context, we will use the term "graphene" for the mixture of mostly FLG with some fraction of SLG. When required the term FLG will be used to emphasize it specific thickness. One should note that, in the considered thickness range, FLG retains its flexibility and remains different from brittle thin films of graphite.

To date, the studies of graphene fillers in TIMs were focused almost exclusively on curing epoxy-based composites. The pioneering studies reported the thermal conductivity enhancement of epoxy by a factor of 25× even at small graphene loading fractions of $f_g = 10$ vol% [42,43]. The only available reports of graphene enhanced non-curing TIMs utilized commercial TIMs with addition of some fraction of graphene fillers. It has been shown that incorporation of small loading fraction graphene fillers into commercial non-curing TIMs enhances their thermal conductivity significantly[42,44–47]. However, in view of undisclosed composition of commercial TIMs it is hard to assess the strength of the effect of graphene fillers. In addition, commercial TIMs already have a high concentrations of fillers, and the addition of even a small amount of graphene results in agglomeration and creation of separated clusters of fillers. These facts motivated the present research, which uses the simple base such as mineral oil and in-house process of preparation and incorporation of graphene fillers.

Combining different types of fillers with various sizes and aspect ratios into a single matrix for achieving the "synergistic effects" is a known strategy for attaining a further enhancement in thermal properties of composites[11,17] . It has been demonstrated that the "synergistic effects" are effective even when one uses fillers of the same material but with two or more size scales[6,17] . A simple explanation for this effect is that smaller size fillers reside between large fillers and connected them more efficiently, leading to improved thermal conduction. By their nature, FLG fillers consist of several graphene monolayers which are held up through weak Van der Waals forces[48] in the cross-plane direction. During the mixing processes of FLG with the matrix materials, due to the high shear stresses involved, the atomic layers of FLG separate out, resulting in a mixture of FLG and SLG fillers, which potentially develop reveal the "synergistic effect". The FLG fillers are better for heat conduction while SLG fillers are more flexible and better for establishing the links among the FLG fillers. These properties can be considered extra advantages of FLG over metallic and ceramic fillers.

In this paper, we describe the details of synthesizing thermal paste with the high loading fraction of FLG fillers with large lateral sizes – in a few μm range. To the best of our knowledge, this is the first report of the non-curing graphene TIMs, prepared without the use of commercial TIM with other fillers. The composition of the non-curing graphene TIM is intentionally kept simple – mineral oil base and FLG fillers. Our results show that even without additional fillers, graphene thermal paste outperforms, in terms of thermal conductivity, the best commercially available non-curing TIMs. The rest of the paper is organized as follows. In Section II, we describe synthesis of graphene thermal paste. Section III provides the results of thermal characterization. The results of benchmarking of graphene paste against the best commercial non-curing TIMs is described in Section IV. Our conclusions are given in Section V.

## II.    Material Synthesis and Characterization

**Sample Preparation:** Figure 1 illustrates the step-by-step preparation procedure and typical applications of non-curing TIMs in electronics. Commercially available graphene fillers (grade H-15, XG-Sciences) with the vendor-specified large lateral dimensions of ~15 µm were weighed and added in pre-calculated proportions to the mineral oil base (Walgreen Health). The large lateral dimensions of the fillers are essential for achieving high thermal conductivity. However, it should be noted that large fillers are more susceptible to rolling and bending during the mixing procedure[3] so special care should be taken in order to avoid filler agglomeration and crumbling, especially at high filler loading fractions. In order to avoid agglomeration, the mixtures of mineral oil and graphene were prepared with addition of acetone in order to keep the filler quality and size intact during the mixing process. Graphene is measured and placed into a container followed by the addition of acetone, creating a graphene-acetone suspension, then the mineral oil is added. The compounds are mixed using a high shear speed mixer (Flacktek Inc.) at 310 rpm, the lowest mixing speed, for about 20 minutes. The effects of mixing speed and other parameters have been researched in details and utilized in the preparation[49,50]. The low mixing speed results in the binding of graphene and mineral oil, and separates them from the acetone which has been added later. The acetone is removed from graphene and mineral oil mixture following phase separation. Finally, the mixture is placed in an oven to evaporate the solvent for ~2 hours at 70° C. This process yields a smooth paste with proper viscosity that is easily spreadable, homogenous, and can be contained within a syringe for later applications. The prepared samples have been characterized using Raman spectroscopy and scanning electron microscopy (SEM) (Supplementary Figure S1 (a-b). The homogenous dispersion of graphene inside the paste is important to the integrity of the composite[51]. In additional to good filler dispersion, a preservation of the fillers throughout the process is another important factor in the performance of the obtained non-curing TIMs.

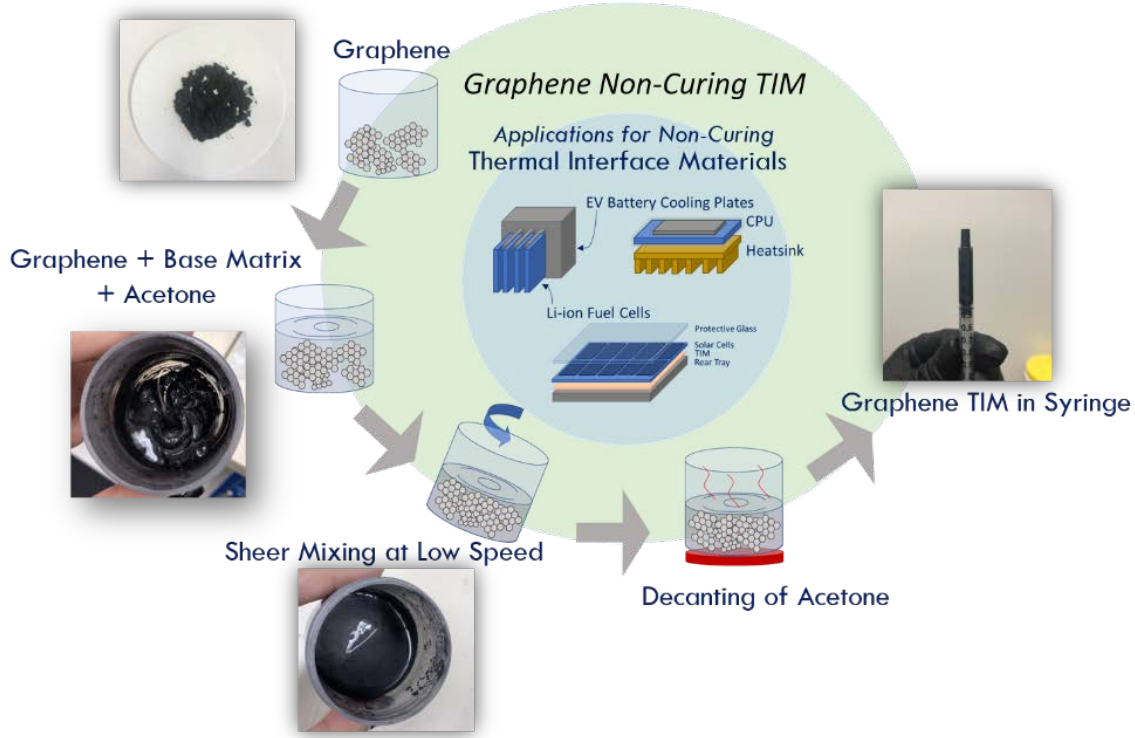

**Figure 1:** Schematic showing typical practical applications of non-curing thermal interface materials in electronics, and the process flow for synthesizing graphene non-curing thermal paste. Graphene is added to the base material with acetone followed by the slow speed sheer mixing. The optimized mixing process seperates the graphene and mineral oil mix from the acetone. This leaves a smooth graphene paste with proper viscosity which is easy to store and apply at the interfaces.

**Thermal Conductivity Measurements and Data Analysis:** The thermal conductivity and contact resistance of the samples were measured using an industrial grade TIM tester (LonGwin Science and Technology Corp.) designed for measurements according to the standard ASTM D 5470-06 – a steady-state method for measuring the thermal properties of TIMs (Supplementary Figure S2). This method is based on the one-dimensional heat conduction Fourier's law, $q'' = -k_{app}\Delta T/\Delta x$, which allows for determining the sample's *apparent* thermal conductivity, $k_{app}$ [Wm$^{-1}$K$^{-1}$], via accurately monitored heat conduction flux, $q''$ [Wm$^{-2}$], and the temperature difference, $\Delta T$ [K], across the sample's thickness, $\Delta x$ [m$^{-1}$]. The sample's total thermal resistance per area, $R''_{tot} = \Delta T/q'' = \frac{\Delta x}{k_{app}}$ [Km$^2$W$^{-1}$], at various thicknesses were measured at a constant temperature of 35 ℃, atmospheric pressure, and plotted as a function of its thickness. The data has been fitted using a linear regression method. The inverse slope and the *y*-intercept of the fitted line shows the TIM's thermal conductivity and twice of its thermal contact resistance, $2R_c''$,[52] as explained below in more details. The temperature dependent thermal conductivity measurements are conducted in the same way, only changing the temperature in the range of 30 ℃ to 115 ℃, with no applied pressure.

### III. Results of Thermal Testing

When a thin layer of TIM is applied between two solid surfaces, assuming a one-dimensional heat flow from the hot to the cold side, the total thermal resistance can be defined as $R_{tot}'' = R_{TIM}'' + R_{c1}'' + R_{c2}''$ where $R_{TIM}''$ is the thermal resistance associated with the TIM layer and $R_{c1}''$ and $R_{c2}''$ are the thermal contact resistances between the TIM and solid surfaces due to the inherent microscopic asperities within solid surfaces (see the schematic in Supplementary Figure S3). This equation can be restated as $R_{tot}'' = BLT/k_{app} = BLT/k_{TIM} + 2R_c''$ considering that the thermal contact resistance between the TIM layer and upper and lower solid surfaces are equal ($R_{c1}'' = R_{c2}'' = R_c''$). In this equation, $k_{app}$ and $k_{TIM}$ are the *apparent* and the "bulk" thermal conductivity of the TIM layer. The difference between these two quantities is that the former depends on BLT and the thermal contact resistances and thus, is not a material property, which is why it is referred as "apparent" thermal conductivity. However, the latter is related to the "true" or "bulk" thermal conductivity of the TIM layer which is a material characteristic and depends on the thermal transport properties of the base polymeric matrix, fillers, and their interaction with each other.

In Figure 2, we show the results of $R_{tot}''$ measurements of TIMs with different filler loadings as a function of the bond line thicknesses (BLT) at a constant temperature of 35 ℃ without any applied pressure (atmospheric pressure). Since $R_{tot}''$ depends linearly on BLT, one can extract $k_{TIM}$ and $R_c''$ with linear fittings (dashed lines in Figure 2) on the experimental data. In this case, the inverse slope and the y-intercept of the fitted line shows $k_{TIM}$ and $2R_c''$, respectively, with an assumption that both parameters remain constant as BLT changes. As one can see, with adding graphene fillers, the slope of the fitted lines decreases significantly, indicating a strong enhancement in the "bulk" thermal conductivity of the compound. However, addition of fillers also increases the thermal contact resistance, which will be discussed later.

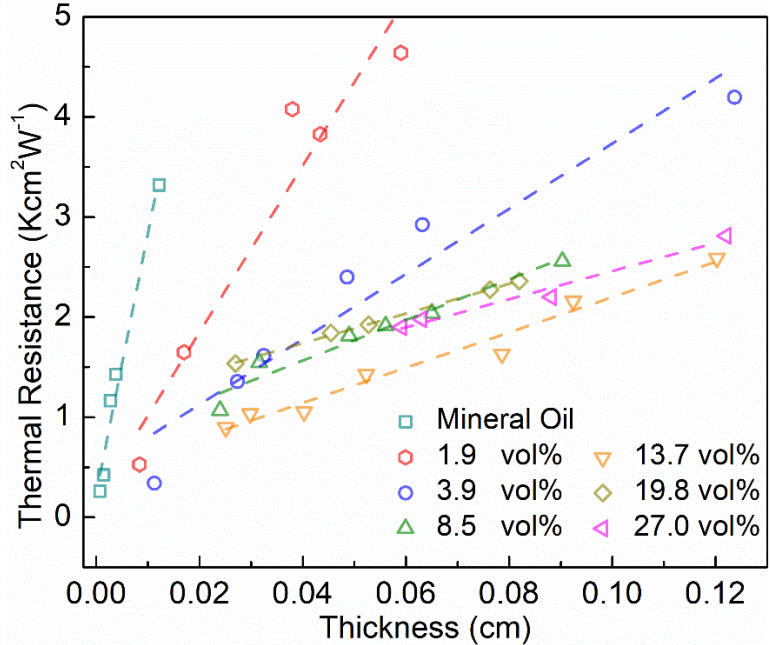

**Figure 2:** Thermal resistance per unit area, $R''$, as a function of the bond line thickness. The dashed lines show the linear regression fittings to the experimental data. Adding graphene fillers to mineral oil results in the slope of the lines decreasing significantly, indicating string enhancement in the "bulk" thermal conductivity of the graphene thermal paste.

Figure 3 presents the thermal conductivity of the graphene non-curing TIMs as a function of the filler loading. The data indicates that at small, $\phi = 1.9$ vol%, graphene filler loading, a significant enhancement in compound's thermal conductivity is observed followed by a saturation effect at the higher loading fractions. This enhancement is attributed to the thermal percolation, i.e. the onset of formation of the continuous network of thermally conductive fillers inside the matrix. The thermal percolation strongly enhances the overall thermal conductivity of the composites. Note that the thermal conductivity increases from 0.3 $Wm^{-1}K^{-1}$ of pure mineral oil to 1.2 $Wm^{-1}K^{-1}$ with addition of only 1.9 vol% of graphene. The observed change in the thermal conductivity is similar to the electrical conductivity behavior of polymers as they are loaded with electrically conductive fillers[53].

In the electrical percolation regime, the electrical conductivity of polymers increases by several orders of magnitude as electrically conductive fillers form a continuous network inside the electrically insulating matrix. The electrical percolation is theoretically described by the power scaling law as $\sigma \sim (\phi - \phi_E)^t$, where $\sigma$ is the electrical conductivity of the composite, $\phi$ is the filler loading fraction, $\phi_E$ is the filler loading at the electrical percolation threshold, and $t$ is the "universal" critical exponent. Following the same theoretical concept, the experimental data in Figure 2 has been fitted by a power scaling law as $k_{TIM} = [(\phi - \phi_{th})/(1 - \phi_{th})]^p$, where $k_f$ is the thermal conductivity of the filler, and $\phi_{th}$ and $p$ are fitting parameters being the filler loading at thermal percolation threshold and the exponent, respectively. The inset in Figure 2 shows the experimental data and theoretical fitting in a log-log scale. Generally, as the loading fraction of filler increases, one would expect to see substantial continuous increases in enhancement of TIM's thermal conductivity. Most cured, *i.e.* solid, TIMs exhibit linear to super-linear thermal conductivity dependence on the filler loading fraction[54]. However, the prepared *non-curing* TIMs exhibit a saturation effect for the thermal conductivity as a function of the filler loading fraction. This is similar to the effect reported previously for nano-fluids and some soft TIMs[5,55–57]. The saturation effect is attributed to a tradeoff between the enhancement in the "bulk" thermal conductivity as more fillers are added to the matrix and the decrease in the thermal conductance as the thermal interface resistance between the filler – filler and filler – matrix interfaces increases due to incorporation of more fillers into the matrix.

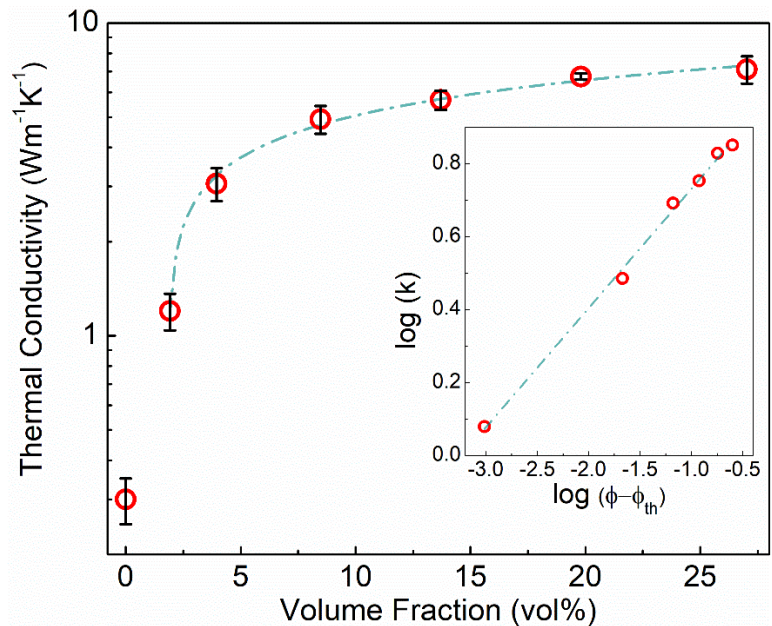

**Figure 3:** Thermal conductivity of the non-curing graphene TIMs as a function of graphene volume fraction. Adding fillers to the mineral oil base leads to more than 4× enhancement of the thermal conductivity at $\phi = 1.9$ vol%. The dashed lines are the theoretical fitting of the experimental data according to the effective thermal conductivity equation $k \sim (\phi - \phi_{th})^p$ with $\phi_{th} = 1.9$ vol% and $p = 0.32$. The inset shows the data in a log-log scale.

Figure 4 shows the contact resistance, $R_c''$, of the non-curing graphene TIMs as a function of the filler loading fraction measured at the atmospheric pressure. As expected, with incorporation of more fillers into the matrix, the contact resistance increases as well. For semi-solid or semi-liquid TIMs, assuming that the "bulk" thermal conductivity of the TIM layer is much smaller than that of the binding surfaces, the contact resistance can be described using the semi-empirical model as $R_{c_1+c_2}'' = 2R_c'' = c\left(\frac{\zeta}{k_{TIM}}\right)\left(\frac{G}{P}\right)^n$ [57] where $G = \sqrt{G'^2 + G''^2}$. In this equation, $G'$ and $G''$ are the storage and loss shear modulus of the TIM, $P$ is the applied pressure, $\zeta$ is the average roughness of the two binding surfaces, assuming that both have the same roughness at interfaces, and $c$ and $n$ are empirical coefficients, respectively. One can see that at a constant applied pressure the prediction of thermal contact resistance becomes cumbersome owing to the fact that the two parameters have opposing effects on the contact resistance. The latter is due to the fact that adding graphene fillers results in increasing both $k_{TIM}$ and $G$. However, this equation intuitively suggests that for TIMs with a specific filler, there exists an optimum filler loading at which the "bulk" thermal conductivity, $k_{TIM}$, significantly increases while the thermal contact resistance, $R_c''$, is affected only slightly. This fact becomes more evident if we restate the total thermal resistance as $R_{tot}'' = \left(\frac{1}{k_{TIM}}\right)\left\{BLT + c\zeta\left(\frac{G}{P}\right)^n\right\}$ showing the importance of increasing the TIM bulk thermal conductivity to reduce the total thermal resistance.

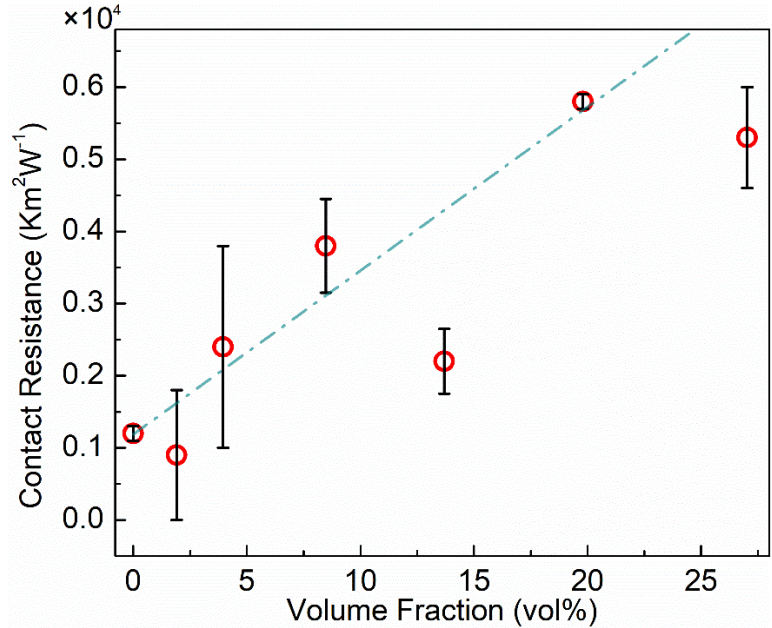

**Figure 4:** Thermal contact resistance as a function of the filler loading. The error bars show the standard error. The thermal contact resistance increases with the loading fraction.

The temperature of electronic devices during operation, no matter how the generated heat is dissipated, increases due to Joule heating, which is unavoidable. The temperature rise depends on the total thermal resistance of the system from the heat source to the environment. In most cases, the TIM layer is one of the bottlenecks for efficient thermal management of the system. In this process, the temperature across the TIM layer increases which in turn, affects its "bulk" thermal conductivity and thermal contact resistance. In order to evaluate the overall thermal performance of the TIM layer in an extended temperature range, the "apparent" thermal conductivity is a more informative parameter. It includes the temperature effects on both the "bulk" thermal conductivity and thermal contact resistance. It is important to evaluate the temperature dependent characteristics of non-curing graphene TIMs in order to verify their overall robustness and stability at elevated temperatures. Practically useful TIMs should perform at high temperatures and retain their thermal properties, as well as sustain an uneven heating throughout the component.

In Figure 5 we present the "apparent" thermal conductivity of the non-curing graphene TIMs as a function of temperature in the range of $40\,°C - 115\,°C$, with no applied pressure. The data are shown for TIMs with various graphene loading fractions. The non-curing graphene TIMs with $\phi = 1.9$ vol% exhibit a slight variation in the "apparent" thermal conductivity as the temperature increases. The "apparent" thermal conductivity change is more pronounced in TIMs with the higher loading compared to that of TIMs with the low graphene loading, although the change is not significant. Generally, the shear modulus of TIMs decreases with increasing the joint temperature, which, in turn, reduces the thermal contact resistance. However, the "bulk" thermal conductivity of TIMs is also decreasing with temperature[45], which causes the overall "apparent" thermal conductivity to drop. At the same time, the decrease in the "apparent" thermal conductivity is not significant, attesting to the practicality of non-curing graphene thermal paste.

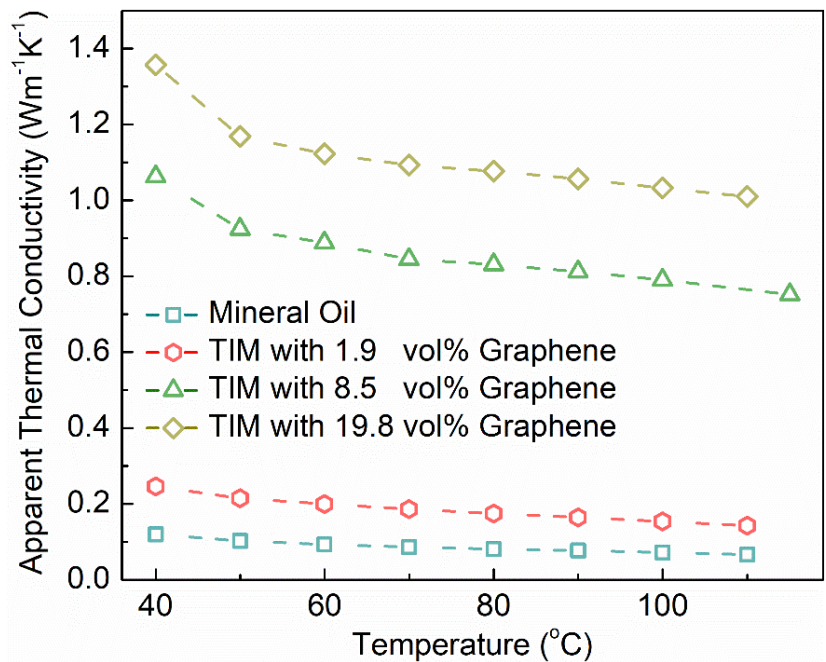

**Figure 5:** "Apparent" thermal conductivity of the non-cured graphene TIMs as a function of temperature in the range from 40 ℃ to 115 ℃.

Another important issue that arises at increased temperature for non-cured TIMs is "bleeding out" [58]. This term indicates the process of thermal grease pumping out from the binding surfaces due to the decrease of the viscosity at elevated temperature[59] . The latter results in reduction of the actual contact of the TIM layer with adjoining solid surfaces, which increases the thermal contact resistance. In order to evaluate the "bleeding out" problem associated with the non-cured graphene TIMs, the BLT variation has been measured as a function of temperature (see Figure 6). The variation in BLT as a function of temperature in pure mineral oil and TIM with $\phi = 1.9$ vol% is

~2 μm/℃ whereas for the TIM with $\phi = 8.5$ vol% it drops to ~0.5 μm/℃. As one can see, the variation is more pronounced at low graphene loadings as compared to that with the high loading. This observation indicates that non-curing graphene TIMs with graphene loading of more than ~8.5 vol% are less prone to the "bleeding out" problem.

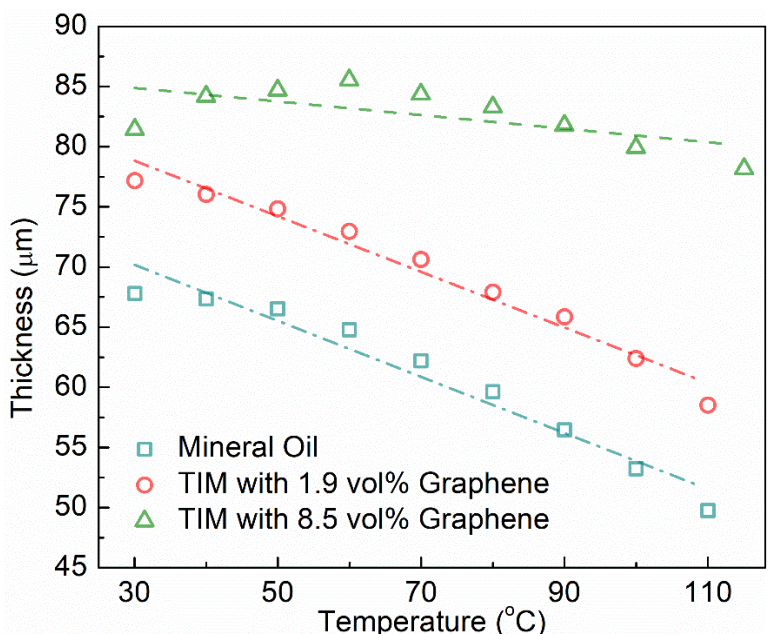

**Figure 6:** Bond line thickness as a function of temperature at atmospheric pressure. The variation in BLT for mineral oil and non-curing graphene TIM with $\phi = 1.9$ vol% is ~2.3 μm/℃. The variation in the TIM thickness with temperature drops to ~0.5 μm/℃ at the higher graphene filler loading. The latter indicates the non-curing graphene TIMs with higher graphene loading are less prone to the "bleeding-out" problem.

## IV. Benchmarking Against Commercial Non-Curing TIMs

In order to benchmark the performance of the non-curing graphene thermal paste against the cutting edge TIM technology, we measured the "bulk" thermal conductivity of five top commercial TIMs widely used in industry. It should be noted that many commercial TIMs claim the thermal conductivity values exceeding $10 \, \mathrm{Wm^{-1}K^{-1}}$, although the vendor supplied descriptions do not specify how the thermal measurements have been performed. In Figure 7, we present the measured and claimed values of the "bulk" thermal conductivity of commercial TIMs. All measurements used the same experimental setup (see above and the Supplemental information) under the same steady-state conditions at 35 ℃ and atmospheric pressure. The obtained data indicate that the true

"bulk" thermal conductivity for all commercial non-curing TIMs is substantially lower than that specified in the vendor datasheets. The thermal conductivity of the non-curing graphene TIM with $\phi = 19.8$ vol% surpasses that of the all commercial TIMs. The highest "bulk" thermal conductivity for the commercial non-curing TIM was obtained for TIM PK-Pro 3 (Prolimatech Inc.). It was determined to be $6.19$ Wm$^{-1}$K$^{-1}$, which is close to the thermal conductivity of the non-curing graphene TIM. However, one should note that PK-Pro 3 incorporates ~90 wt% of Al and ZnO as fillers[60] while graphene TIMs includes only 40 wt% of graphene (see Figure 7).

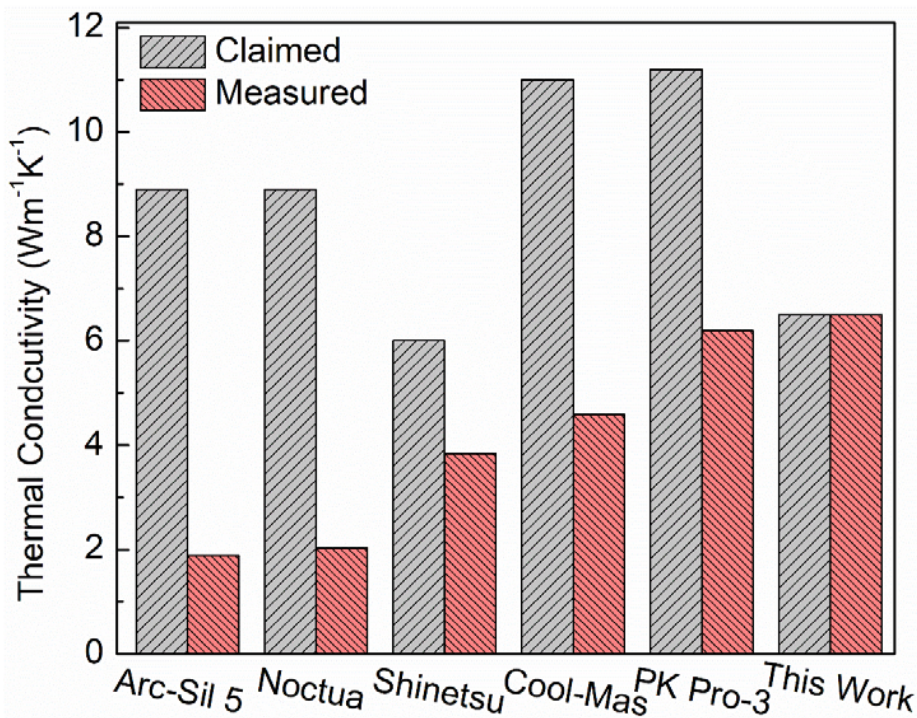

**Figure 7:** Benchmarking of non-curing graphene TIMs against commercial non-curing TIMs. The non-curing graphene TIM has $\phi = 19.8$ vol% (40 wt%) filler loading. The grey bars show the thermal conductivity values claimed by the vendors. The light coral bars present the data measured by the same instrument used for this study. There is a substantial discrepancy between the claimed and measured data for the commercial TIMs. The non-curing graphene thermal paste outperforms all commercial non-curing TIMs. A commercial non-curing TIM with the highest thermal conductivity (PK Pro-3) uses ~90 wt% of Al and ZnO filler loading, which is more than two times of the graphene filler concentration used in this study.

Table I summarizes recent research data for non-curing TIMs and nano-fluids with different fillers and host matrices. The difficulty in uniform dispersion of fillers through the matrix could be one of the reasons for the scarcity of literature in the field of non-curing thermal interface materials.

The data presented in Figure 7 and Table I attest for the great potential of non-curing graphene thermal paste for thermal management of advanced electronics.

**Table I:** Thermal conductivity of non-curing thermal interface materials with different fillers

| Filler | Base Matrix | Filler Loading vol.% | wt % | Method | K ($Wm^{-1}K^{-1}$) | Refs. |
|---|---|---|---|---|---|---|
| Graphene | Mineral Oil | 27 | 50 | Time tester | 7.1 | This work |
| $Al_2O_3$ / Graphene | Silicone grease | 6/1 | - | TPS | 3.0 | [17] |
| Graphene | Epoxy without resin | 11 | - | Tim tester | 0.90 | [61] |
| rGO | Silicon Oil | 4.3 | - | THWM | 1 | [62] |
| Graphene NF | Silicon Oil | 4.3 | - | THWM | 0.25 | [62] |
| Graphene | Silicon Oil | 0.07 | - | THWM | 0.215 | [63] |
| Graphene / CuO | Water | 0.07 | - | Kd2 thermometer | 0.28 | [64] |
| Graphene / $Fe_3O_4$ | Commercial TIM | - | 6 | Laser flash | ~1.46 | [46] |
| Functionalized Graphene | Water | - | 5 | THWM | 1.15 | [65] |
| GNP | Silicone Grease | 0.75 | - | THB | 3.2 | [66] |
| GNP | Water | - | 0.10 | THWM | 0.75 | [67] |
| CNT | Silicone Elastomer | - | 4 | Tim tester | 1.8 | [18] |
| Silica | water | 3 | - | THWM | 0.66 | [68] |
| CuO microdisks | Silicone Base | 0.09 | - | Hot disk | 0.28 | [20] |
| CuO nanoblock | Silicone Base | 0.09 | - | Hot disk | 0.25 | [20] |
| CuO microspheres | Silicone Base | 0.09 | - | Hot disk | 0.23 | [69] |
| $TiO_2$ | Water | 5 | - | THWM | 0.871 | [15] |
| AlN | EG, PG | 10 | - | THWM | 0.35 | [13] |
| T- ZnO | Silicon Oil | 18 | - | TPS | 0.88 | [19] |
| NF = nano-flakes, GNP = graphene nano-platelets, EG = Ethylene Glycol, PG = Propylene Glycol THWM = transient hot wire method, TPS = Transient plane source THB = transient hot bridge | | | | | | |

## V. Conclusions

We reported on the synthesis and thermal conductivity measurements of non-curing thermal paste, i.e. grease, based on mineral oil with the mixture of graphene and few-layer graphene flakes as the fillers. It was found that graphene thermal paste exhibits a distinctive thermal percolation threshold with the thermal conductivity revealing a sublinear dependence on the filler loading. This behavior contrasts with the thermal conductivity of curing graphene thermal interface materials, based on epoxy, where super-linear dependence on the filler loading is observed. The performance of graphene thermal paste was benchmarked against top-of-the-line commercial thermal pastes. The obtained results show that non-curing graphene thermal interface materials outperforms the best commercial pastes in terms of thermal conductivity, at substantially lower filler concentration. The obtained results shed light on thermal percolation mechanism in non-curing polymeric matrices laden with quasi-two-dimensional fillers. Considering recent progress in graphene production via liquid phase exfoliation and oxide reduction, we argue that our results open a pathway for large-scale industrial application of graphene in thermal management of electronics.

## Supporting Information

Supporting Information is available from the journal web-site for free of charge.

## Acknowledgements

This work was supported, in part, by the UC – National Laboratory Collaborative Research and Training Program.

## Contributions

A.A.B. and F.K. conceived the idea of the study and coordinated the project. S.N. prepared the thermal compounds, performed sample characterization, thermal conductivity measurements, and conducted the data analysis; F.K. coordinated the project and contributed to the experimental and theoretical data analysis; D.W. assisted in preparation of the samples, measurements of the commercial materials; C.T.H performed scanning electron microscopy of the samples; A.M. measured Raman spectra of the samples; Z.B. helped in the sample preparation; R.S. contributed to the analysis of experimental data; A.A.B. coordinated the project and contributed to the thermal data analysis; F.K. and A.A.B. led the manuscript preparation. All authors contributed to writing and editing of the manuscript.

**References**

(1)     Smoyer, J. L.; Norris, P. M. Brief Historical Perspective in Thermal Management and the Shift Toward Management at the Nanoscale. *Heat Transf. Eng.* **2019**, *40* (3–4), 269–282. https://doi.org/10.1080/01457632.2018.1426265.

(2)     Kargar, F.; Barani, Z.; Balinskiy, M.; Magana, A. S.; Lewis, J. S.; Balandin, A. A. Dual-Functional Graphene Composites for Electromagnetic Shielding and Thermal Management. *Adv. Electron. Mater.* **2018**, *0* (0), 1800558. https://doi.org/10.1002/aelm.201800558.

(3)     Kargar, F.; Barani, Z.; Salgado, R.; Debnath, B.; Lewis, J. S.; Aytan, E.; Lake, R. K.; Balandin, A. A. Thermal Percolation Threshold and Thermal Properties of Composites with High Loading of Graphene and Boron Nitride Fillers. *ACS Appl. Mater. Interfaces* **2018**, *10* (43), 37555–37565. https://doi.org/10.1021/acsami.8b16616.

(4)     Cola, B. a.; Xu, X.; Fisher, T. S. Increased Real Contact in Thermal Interfaces: A Carbon Nanotube/Foil Material. *Appl. Phys. Lett.* **2007**, *90* (9), 88–91. https://doi.org/10.1063/1.2644018.

(5)     Mu, L.; Ji, T.; Chen, L.; Mehra, N.; Shi, Y.; Zhu, J. Paving the Thermal Highway with Self-Organized Nanocrystals in Transparent Polymer Composites. **2016**. https://doi.org/10.1021/acsami.6b10451.

(6)     Lewis, J. S.; Barani, Z.; Magana, A. S.; Kargar, F.; Balandin, A. A. Thermal and Electrical Conductivity Control in Hybrid Composites with Graphene and Boron Nitride Fillers. *Mater. Res. Express* **2019**, *6* (8). https://doi.org/10.1088/2053-1591/ab2215.

(7)     Teng, C.-C.; Ma, C.-C. M.; Lu, C.-H.; Yang, S.-Y.; Lee, S.-H.; Hsiao, M.-C.; Yen, M.-Y.; Chiou, K.-C.; Lee, T.-M. Thermal Conductivity and Structure of Non-Covalent

Functionalized Graphene/Epoxy Composites. *Carbon N. Y.* **2011**, *49* (15), 5107–5116. https://doi.org/10.1016/J.CARBON.2011.06.095.

(8)     Data     Centers     and     Servers     |     Department     of     Energy https://www.energy.gov/eere/buildings/data-centers-and-servers (accessed Feb 11, 2019).

(9)     Gwinn, J. P.; Webb, R. L. Performance and Testing of Thermal Interface Materials. https://doi.org/10.1016/S0026-2692(02)00191-X.

(10)   Bar-Cohen, A.; Matin, K.; Narumanchi, S. Nanothermal Interface Materials: Technology Review and Recent Results. *J. Electron. Packag.* **2015**, *137* (4), 040803. https://doi.org/10.1115/1.4031602.

(11)   Barani, Z.; Mohammadzadeh, A.; Geremew, A.; Huang, C.; Coleman, D.; Mangolini, L.; Kargar, F.; Balandin, A. A. Thermal Properties of the Binary-Filler Hybrid Composites with Graphene   and   Copper   Nanoparticles. *Adv.   Funct.   Mater.* **2019**, 1904008. https://doi.org/10.1002/adfm.201904008.

(12)   Xie, X.; Li, D.; Tsai, T.-H.; Liu, J.; Braun, P. V.; Cahill, D. G. Thermal Conductivity, Heat Capacity, and Elastic Constants of Water-Soluble Polymers and Polymer Blends. *Macromolecules* **2016**, *49* (3), 972–978. https://doi.org/10.1021/acs.macromol.5b02477.

(13)   Yu, W.; Xie, H.; Li, Y.; Chen, L. Experimental Investigation on Thermal Conductivity and Viscosity of Aluminum Nitride Nanofluid. *Particuology* **2011**, *9* (2), 187–191. https://doi.org/10.1016/J.PARTIC.2010.05.014.

(14)   Zeng, J. L.; Cao, Z.; Yang, D. W.; Sun, L. X.; Zhang, L. Thermal Conductivity Enhancement of Ag Nanowires on an Organic Phase Change Material. In *Journal of Thermal   Analysis   and   Calorimetry*;   2010;   Vol.   101,   pp   385–389.

https://doi.org/10.1007/s10973-009-0472-y.

(15)   Murshed, S. M. S.; Leong, K. C.; Yang, C. Enhanced Thermal Conductivity of TiO2—Water Based Nanofluids. *Int. J. Therm. Sci.* **2005**, *44* (4), 367–373. https://doi.org/10.1016/J.IJTHERMALSCI.2004.12.005.

(16)   Wong, C. P.; Bollampally, R. S. Thermal Conductivity, Elastic Modulus, and Coefficient of Thermal Expansion of Polymer Composites Filled with Ceramic Particles for Electronic Packaging. *J. Appl. Polym. Sci.* **1999**, *74* (14), 3396–3403. https://doi.org/10.1002/(SICI)1097-4628(19991227)74:14<3396::AID-APP13>3.0.CO;2-3.

(17)   Yu, W.; Xie, H.; Yin, L.; Zhao, J.; Xia, L.; Chen, L. Exceptionally High Thermal Conductivity of Thermal Grease: Synergistic Effects of Graphene and Alumina. *Int. J. Therm. Sci.* **2015**, *91*, 76–82. https://doi.org/10.1016/J.IJTHERMALSCI.2015.01.006.

(18)   Chen, H.; Wei, H.; Chen, M.; Meng, F.; Li, H.; Li, Q. Enhancing the Effectiveness of Silicone Thermal Grease by the Addition of Functionalized Carbon Nanotubes. *Appl. Surf. Sci.* **2013**, *283*, 525–531. https://doi.org/10.1016/J.APSUSC.2013.06.139.

(19)   Du, H.; Qi, Y.; Yu, W.; Yin, J.; Xie, H. T-Shape ZnO Whisker: A More Effective Thermal Conductive Filler than Spherical Particles for the Thermal Grease. *Int. J. Heat Mass Transf.* **2017**, *112*, 1052–1056. https://doi.org/10.1016/j.ijheatmasstransfer.2017.05.016.

(20)   Yu, W.; Zhao, J.; Wang, M.; Hu, Y.; Chen, L.; Xie, H. Thermal Conductivity Enhancement in Thermal Grease Containing Different CuO Structures. *Nanoscale Res. Lett.* **2015**, *10*, 113. https://doi.org/10.1186/s11671-015-0822-6.

(21)   Cola, B. A.; Xu, X.; Fisher, T. S. Increased Real Contact in Thermal Interfaces: A Carbon

Nanotube/Foil Material. *Appl. Phys. Lett.* **2007**, *90* (9), 093513. https://doi.org/10.1063/1.2644018.

(22) Initial hype passed, graphene shows up in real products https://www.canadianbusiness.com/innovation/graphene-after-the-hype/ (accessed Oct 10, 2019).

(23) Graphene thermal conductivity - introduction and latest news | Graphene-Info https://www.graphene-info.com/graphene-thermal (accessed Oct 10, 2019).

(24) Novoselov, K. S.; Geim, A. K.; Morozov, S. V; Jiang, D.; Zhang, Y.; Dubonos, S. V; Grigorieva, I. V; Firsov, A. A. Electric Field Effect in Atomically Thin Carbon Films. *Science (80-. ).* **2004**, *306* (5696), 666 LP – 669.

(25) Geim, A. K.; Novoselov, K. S. The Rise of Graphene. *Nat. Mater.* **2007**, *6* (3), 183–191. https://doi.org/10.1038/nmat1849.

(26) Shao, Q.; Liu, G.; Teweldebrhan, D.; Balandin, A. A. Higherature Quenching of Electrical Resistance in Graphene Interconnects. *Appl. Phys. Lett.* **2008**, *92* (19). https://doi.org/10.1063/1.2927371.

(27) Schwierz, F. Graphene Transistors. *Nat. Nanotechnol.* **2010**, *5* (7), 487–496. https://doi.org/10.1038/nnano.2010.89.

(28) Yang, X.; Liu, G.; Balandin, A. A.; Mohanram, K. Triple-Mode Single-Transistor Graphene Amplifier and Its Applications. *ACS Nano* **2010**, *4* (10), 5532–5538. https://doi.org/10.1021/nn1021583.

(29) Yang, X.; Liu, G.; Rostami, M.; Balandin, A. A.; Mohanram, K. Graphene Ambipolar Multiplier Phase Detector. *IEEE Electron Device Lett.* **2011**, *32* (10), 1328–1330.

https://doi.org/10.1109/LED.2011.2162576.

(30)  Liu, G.; Ahsan, S.; Khitun, A. G.; Lake, R. K.; Balandin, A. A. Graphene-Based Non-Boolean Logic Circuits. *J. Appl. Phys.* **2013**, *114* (15). https://doi.org/10.1063/1.4824828.

(31)  Balandin, A. A. Thermal Properties of Graphene and Nanostructured Carbon Materials. *Nature Materials*. Nature Publishing Group August 1, 2011, pp 569–581. https://doi.org/10.1038/nmat3064.

(32)  Balandin, A. A.; Ghosh, S.; Bao, W.; Calizo, I.; Teweldebrhan, D.; Miao, F.; Lau, C. N. Superior Thermal Conductivity of Single-Layer Graphene. *Nano Lett.* **2008**, *8* (3), 902–907. https://doi.org/10.1021/nl0731872.

(33)  Li, H.; Ying, H.; Chen, X.; Nika, D. L.; Cocemasov, A. I.; Cai, W.; Balandin, A. A.; Chen, S. Thermal Conductivity of Twisted Bilayer Graphene. *Nanoscale* **2014**, *6* (22), 13402–13408. https://doi.org/10.1039/C4NR04455J.

(34)  Nika, D. L.; Pokatilov, E. P.; Askerov, A. S.; Balandin, A. A. Phonon Thermal Conduction in Graphene: Role of Umklapp and Edge Roughness Scattering. https://doi.org/10.1103/PhysRevB.79.155413.

(35)  Nika, D. L.; Ghosh, S.; Pokatilov, E. P.; Balandin, A. A. Lattice Thermal Conductivity of Graphene Flakes: Comparison with Bulk Graphite. *Appl. Phys. Lett.* **2009**, *94* (20), 203103. https://doi.org/10.1063/1.3136860.

(36)  Nika, D. L.; Askerov, A. S.; Balandin, A. A. Anomalous Size Dependence of the Thermal Conductivity of Graphene Ribbons. *Nano Lett.* **2012**, *12* (6), 3238–3244. https://doi.org/10.1021/nl301230g.

(37)  Hernandez, Y.; Nicolosi, V.; Lotya, M.; Blighe, F. M.; Sun, Z.; De, S.; McGovern, I. T.;

Holland, B.; Byrne, M.; Gun'Ko, Y. K.; et al. High-Yield Production of Graphene by Liquid-Phase Exfoliation of Graphite. *Nat. Nanotechnol.* **2008**, *3* (9), 563–568. https://doi.org/10.1038/nnano.2008.215.

(38)    Lotya, M.; Hernandez, Y.; King, P. J.; Smith, R. J.; Nicolosi, V.; Karlsson, L. S.; Blighe, F. M.; De, S.; Wang, Z.; McGovern, I. T.; et al. Liquid Phase Production of Graphene by Exfoliation of Graphite in Surfactant/Water Solutions. *J. Am. Chem. Soc.* **2009**, *131* (10), 3611–3620. https://doi.org/10.1021/ja807449u.

(39)    Stankovich, S.; Dikin, D. A.; Piner, R. D.; Kohlhaas, K. A.; Kleinhammes, A.; Jia, Y.; Wu, Y.; Nguyen, S. T.; Ruoff, R. S. Synthesis of Graphene-Based Nanosheets via Chemical Reduction of Exfoliated Graphite Oxide. *Carbon N. Y.* **2007**, *45* (7), 1558–1565. https://doi.org/10.1016/j.carbon.2007.02.034.

(40)    Stankovich, S.; Dikin, D. A.; Dommett, G. H. B.; Kohlhaas, K. M.; Zimney, E. J.; Stach, E. A.; Piner, R. D.; Nguyen, S. T.; Ruoff, R. S. Graphene-Based Composite Materials. **2006**, *442* (20). https://doi.org/10.1038/nature04969.

(41)    Nethravathi, C.; Rajamathi, M. Chemically Modified Graphene Sheets Produced by the Solvothermal Reduction of Colloidal Dispersions of Graphite Oxide. *Carbon N. Y.* **2008**, *46* (14), 1994–1998. https://doi.org/10.1016/j.carbon.2008.08.013.

(42)    Shahil, K. M. F.; Balandin, A. A. Graphene–Multilayer Graphene Nanocomposites as Highly Efficient Thermal Interface Materials. *Nano Lett.* **2012**, *12* (2), 861–867. https://doi.org/10.1021/nl203906r.

(43)    Shahil, K. M. F.; Balandin, A. A. Thermal Properties of Graphene and Multilayer Graphene: Applications in Thermal Interface Materials. *Solid State Commun.* **2012**, *152* (15), 1331–

1340. https://doi.org/10.1016/J.SSC.2012.04.034.

(44)  Hansson, J.; Zandén, C.; Ye, L.; Liu, J. *Review of Current Progress of Thermal Interface Materials for Electronics Thermal Management Applications*.

(45)  Goyal, V.; Balandin, A. A. Thermal Properties of the Hybrid Graphene-Metal Nano-Micro-Composites: Applications in Thermal Interface Materials. *Appl. Phys. Lett.* **2012**. https://doi.org/10.1063/1.3687173.

(46)  Renteria, J.; Legedza, S.; Salgado, R.; Balandin, M. P.; Ramirez, S.; Saadah, M.; Kargar, F.; Balandin, A. A. Magnetically-Functionalized Self-Aligning Graphene Fillers for High-Efficiency Thermal Management Applications. *Mater. Des.* **2015**, *88*, 214–221. https://doi.org/10.1016/j.matdes.2015.08.135.

(47)  Kargar, F.; Salgado, R.; Legedza, S.; Renteria, J.; Balandin, A. A. A Comparative Study of the Thermal Interface Materials with Graphene and Boron Nitride Fillers; Razeghi, M., Lee, Y. H., Ghazinejad, M., Eds.; International Society for Optics and Photonics, 2014; Vol. 9168, p 91680S. https://doi.org/10.1117/12.2070704.

(48)  Nair, R. R.; Blake, P.; Grigorenko, A. N.; Novoselov, K. S.; Booth, T. J.; Stauber, T.; Peres, N. M. R.; Geim, A. K. Fine Structure Constant Defines Visual Transparency of Graphene. *Science (80-. ).* **2008**, *320* (5881), 1308. https://doi.org/10.1126/science.1156965.

(49)  Arao, Y.; Mizuno, Y.; Araki, K.; Kubouchi, M. Mass Production of High-Aspect-Ratio Few-Layer-Graphene by High-Speed Laminar Flow. *Carbon N. Y.* **2016**, *102*, 330–338. https://doi.org/10.1016/J.CARBON.2016.02.046.

(50)  Aiping Yu; Palanisamy Ramesh; Mikhail E. Itkis; Elena Bekyarova,  and; Haddon*, R. C. Graphite Nanoplatelet−Epoxy Composite Thermal Interface Materials. **2007**.

https://doi.org/10.1021/JP071761S.

(51)    Xu, X.; Pereira, L. F. C.; Wang, Y.; Wu, J.; Zhang, K.; Zhao, X.; Bae, S.; Tinh Bui, C.; Xie, R.; Thong, J. T. L.; et al. Length-Dependent Thermal Conductivity in Suspended Single-Layer Graphene. *Nat. Commun.* **2014**, *5* (1), 3689. https://doi.org/10.1038/ncomms4689.

(52)    *Characteristics of Thermal Interface Materials*; 2001.

(53)    Bonnet, P.; Sireude, D.; Garnier, B.; Chauvet, O. Thermal Properties and Percolation in Carbon Nanotube-Polymer Composites. *Appl. Phys. Lett.* **2007**, *91* (20). https://doi.org/10.1063/1.2813625.

(54)    Kargar, F.; Barani, Z.; Balinskiy, M.; Magana, A. S.; Lewis, J. S.; Balandin, A. A.; Salgado, R.; Debnath, B.; Lewis, J. S.; Aytan, E.; et al. Dual-Functional Graphene Composites for Electromagnetic Shielding and Thermal Management. *Adv. Electron. Mater.* **2018**, *10* (43), 37555–37565. https://doi.org/10.1002/aelm.201800558.

(55)    Zhang, L.; Ruesch, M.; Zhang, X.; Bai, Z.; Liu, L. Tuning Thermal Conductivity of Crystalline Polymer Nanofibers by Interchain Hydrogen Bonding †. **2015**. https://doi.org/10.1039/c5ra18519j.

(56)    Evans, W.; Prasher, R.; Fish, J.; Meakin, P.; Phelan, P.; Keblinski, P. Effect of Aggregation and Interfacial Thermal Resistance on Thermal Conductivity of Nanocomposites and Colloidal Nanofluids. **2007**. https://doi.org/10.1016/j.ijheatmasstransfer.2007.10.017.

(57)    Prasher, R. S.; Shipley, J.; Prstic, S.; Koning, P.; Wang, J.-L. *THERMAL RESISTANCE OF PARTICLE LADEN POLYMERIC THERMAL INTERFACE MATERIALS*; 2003.

(58)    Skuriat, R.; Li, J. F.; Agyakwa, P. A.; Mattey, N.; Evans, P.; Johnson, C. M. Degradation of Thermal Interface Materials for High-Temperature Power Electronics Applications.

*Microelectron. Reliab.* **2013**, *53* (12), 1933–1942. https://doi.org/10.1016/j.microrel.2013.05.011.

(59)   Gao, J.; Mwasame, P. M.; Wagner, N. J. Thermal Rheology and Microstructure of Shear Thickening Suspensions of Silica Nanoparticles Dispersed in the Ionic Liquid [C 4 Mim][BF 4 ] . *J. Rheol. (N. Y. N. Y).* **2017**, *61* (3), 525–535. https://doi.org/10.1122/1.4979685.

(60)   Prolimatech http://www.prolimatech.com/en/ (accessed Nov 20, 2019).

(61)   Tian, X.; Itkis, M. E.; Bekyarova, E. B.; Haddon, R. C. Anisotropic Thermal and Electrical Properties of Thin Thermal Interface Layers of Graphite Nanoplatelet-Based Composites. *Sci. Rep.* **2013**, *3* (1), 1710. https://doi.org/10.1038/srep01710.

(62)   Yu, W.; Xie, H.; Chen, L.; Zhu, Z.; Zhao, J.; Zhang, Z. Graphene Based Silicone Thermal Greases. *Phys. Lett. A* **2014**, *378* (3), 207–211. https://doi.org/10.1016/J.PHYSLETA.2013.10.017.

(63)   Ma, W.; Yang, F.; Shi, J.; Wang, F.; Zhang, Z.; Wang, S. Silicone Based Nanofluids Containing Functionalized Graphene Nanosheets. *Colloids Surfaces A Physicochem. Eng. Asp.* **2013**, *431*, 120–126. https://doi.org/10.1016/J.COLSURFA.2013.04.031.

(64)   Baby, T. T.; Sundara, R. Synthesis and Transport Properties of Metal Oxide Decorated Graphene Dispersed Nanofluids. *J. Phys. Chem. C* **2011**, *115* (17), 8527–8533. https://doi.org/10.1021/jp200273g.

(65)   Ghozatloo, A.; Shariaty-Niasar, M.; Rashidi, A. M. Preparation of Nanofluids from Functionalized Graphene by New Alkaline Method and Study on the Thermal Conductivity and Stability. *Int. Commun. Heat Mass Transf.* **2013**, *42*, 89–94.

https://doi.org/10.1016/J.ICHEATMASSTRANSFER.2012.12.007.

(66)  Phuong, M.T.; Trinh, P.V.; Tuyen, N.V.; Dinh, N.N.; Minh, P.N.; Dung, N.D.; Thang, B. H. Effect of Graphene Nanoplatelet Concentration on the Thermal Conductivity of Silicone Thermal Grease. *Physics, J. Nano- Electron.* **2019**, *11* (5), 5039. https://doi.org/10.21272/jnep.11(5).05039.

(67)  Mehrali, M.; Sadeghinezhad, E.; Latibari, S. T.; Kazi, S. N.; Mehrali, M.; Bin, M. N.; Zubir, M.; Simon, H.; Metselaar, C. *Investigation of Thermal Conductivity and Rheological Properties of Nanofluids Containing Graphene Nanoplatelets*; 2014.

(68)  Ranjbarzadeh, R.; Moradikazerouni, A.; Bakhtiari, R.; Asadi, A.; Afrand, M. An Experimental Study on Stability and Thermal Conductivity of Water/Silica Nanofluid: Eco-Friendly Production of Nanoparticles. *J. Clean. Prod.* **2019**, *206*, 1089–1100. https://doi.org/10.1016/j.jclepro.2018.09.205.

(69)  Yu, W.; Zhao, J.; Wang, M.; Hu, Y.; Chen, L.; Xie, H. Thermal Conductivity Enhancement in Thermal Grease Containing Different CuO Structures. *Nanoscale Res. Lett.* **2015**, *10* (1), 113. https://doi.org/10.1186/s11671-015-0822-6.